\documentclass[printer]{tMOP2e}

\newcommand{\eg}{e.g.~}
\newcommand{\ie}{i.e.~}
\newcommand{\var}{\mathrm{var}}
\newcommand{\rmi}{\mathrm{i}}
\newcommand{\erf}{\mathrm{erf}}
\newcommand{\Real}{\mathrm{Re}}

\title{Angular EPR paradox}
\author{\uppercase{J. B. G{\"o}tte, S. Franke-Arnold and Stephen M. Barnett} \\
Department of Physics, University of Strathclyde, Glasgow G4 0NG, United Kingdom}
\received{}

\jname{Journal of Modern Optics}
\jvol{00}
\jnum{00}
\jmonth{00}
\jyear{00}

\begin{document}

\maketitle

%%% PREAMBLE

%%% ABSTRACT

\begin{abstract}
The violation of local uncertainty relations is a valuable tool for detecting 
entanglement, especially in multi-dimensional systems. The orbital angular momentum of light provides such a 
multi-dimensional system. We study quantum correlations for the conjugate variables of orbital angular momentum 
and angular position. We determine an experimentally testable criterion for the demonstration of an angular version 
of the EPR paradox. For the interpretation of future experimental results 
from our proposed setup, we include a model for the indeterminacies inherent to the angular position measurement.
For this measurement angular apertures are used to determine the probability density of the angle.
We show that for a class of aperture functions a demonstration of an angular EPR paradox, according to our criterion, 
is to be expected. 
\end{abstract}

%%% MAIN DOCUMENT

%----------------------------

%%% SECTION: Introduction
\section{Introduction} 
Experiments on the orbital angular momentum [OAM] of light confirmed recently an uncertainty
principle for angular position and angular momentum \cite{frankearnold+:njp6:2004}. Whereas for
separable quantum states
uncertainty principles limit the accuracy for measurements of non-commuting observables, 
inseparable or entangled states may apparently overcome these limits.
This was first discussed by Einstein, Podolsky and Rosen in their famous \textit{Gedankenexperiment}
\cite{EPR:pr47:1935} and led to the formulation of the EPR paradox \cite{bohm:ph:1951}.
The implications of the EPR paradox have been tested mainly on optical systems, \eg on the polarisation 
of photons \cite{aspect+:prl49:1982,weihs+:prl81:1998}, quadrature phase components 
\cite{reid:pra40:1989,ou+:prl68:1992} or directly on the optical version of EPR's original example, 
the linear momentum and linear position of photons \cite{howell+:prl92:2004}. The relation 
between OAM and its conjugate variable, the angular position, is fundamentally different from these 
systems, because OAM is a discrete quantum observable of infinite dimension and the angular position is continuous
and bounded. The entanglement for OAM of photon pairs generated in parametric down conversion has been confirmed 
both experimentally \cite{mair+:nat412:2001} and theoretically \cite{frankearnold+:pra65:2002}. 
It is therefore interesting 
to examine the possibility to demonstrate an `angular' EPR paradox for this pair of observables, in particular as
the necessary experimental techniques have already been employed in recent work in this field 
\cite{frankearnold+:njp6:2004,leach+:prl88:2002}.

Apart from the fundamental interest in angular EPR correlations, the criterion for 
an EPR paradox for OAM and angular position of light provides a tool to characterise
entanglement for these observables. Using variances in violations of local uncertainty relations
as entanglement criteria has received renewed interest. Formerly applied to continuous
variables on a specific example \cite{reid:pra40:1989}, it has been shown that variances of special observables 
can be used to detect entanglement in finite-dimensional system \cite{hofmann+:pra68:2003}. This approach has been
generalised to arbitrary observables and led to entanglement criteria for finite dimensional systems 
\cite{guehne:prl92:2004} which are known from the continuous variable regime \cite{werner:prl86:2001}.

A practical motivation for studying entanglement of optical OAM arises 
from the possible advantages of OAM in quantum information processes \cite{molinaterriza+:prl88:2002} 
and quantum communication, as cryptographic schemes \cite{ekert:prl67:1991} could profit from 
an enlarged basis of states.

%%%%% SUBSECTION: Orbital angular momentum

\subsection{Orbital angular momentum}
\label{ssc:oam}
OAM of light is connected with the azimuthal phase structure of light beams: each photon in a beam with phase dependence
$\exp(\rmi m \phi)$ carries an OAM of $m\hbar$ \cite{allen+:pra45:1992,allen+:pro39:1999,allen+:iop:2003}. The OAM 
number $m \in \mathbf{Z}$ can take on 
any integer number which leads to a discrete, infinite dimensional quantum system. The conjugate variable is the angular 
or azimuthal position $\phi_\theta \in [\theta, 2\pi + \theta)$, which we choose to lie within the $2\pi$ radian interval
starting at angle $\theta$. The associated uncertainty relation is then \cite{barnettpegg:pra41:1990,frankearnold+:njp6:2004}
\begin{equation}
\label{eq:oamur}
  \Delta L_z \Delta \phi_\theta \geq \frac{1}{2} \hbar | 1 - 2\pi P(\theta) |,
\end{equation}
where $\Delta L_z = \hbar \Delta m$ and $P(\theta)$ is the angular probability density at the boundary of the 
chosen interval. The topology of the basis sets is reflected in their Fourier relation; in contrast to the linear 
case a discrete Fourier transform allows us to change from OAM representation to angle representation:
\begin{eqnarray}
\langle\phi | \psi \rangle = \psi(\phi) = \frac{1}{\sqrt{2\pi}} \sum_{m \in \mathbf{Z}} \exp(-\rmi m \phi) c_m, \\
\label{eq:fouriertrans}	
\langle m | \psi \rangle = c_m = \frac{1}{\sqrt{2\pi}} \int_\theta^{2\pi + \theta} d\phi \exp(\rmi m \phi) \psi(\phi).
\end{eqnarray}
Here, $\psi(\phi)$ is the wavefunction in the angle representation and $c_m$ the OAM probability amplitude. 
The $2\pi$ radian interval is commonly chosen to be $[-\pi,\pi)$, and we set $\theta = -\pi$ from here on.
 
It has been experimentally demonstrated that OAM of light is conserved under parametric down conversion \cite{mair+:nat412:2001}. 
In theoretical
studies, it has been pointed out that the conservation of OAM is related to the phase matching condition for parametric down conversion 
processes \cite{frankearnold+:pra65:2002}. Entanglement in OAM and azimuthal position is a consequence of this phase matching 
\cite{allen+:iop:2003}. 
The conservation of transverse momentum requires that the two-photon wavefunction for the signal (1) and idler (2) mode,
for a plane wave pump, has to be of the form
$\delta(k_{1,x} + k_{2_x})\delta(k_{1,y} + k_{2,y})$. Using a simplified approach one can argue that the transverse spatial 
correlations in the far field originate from the momentum conservation under parametric down conversion. Identifying transverse 
momentum components in the near field with spatial coordinates in the far field allows us to write the spatial dependence of the 
wavefunction in position representation as
\begin{equation}
\label{eq:anglecorr}	
\delta(x_1 + x_2)\delta(y_1 + y_2) = \frac{1}{\rho_1}\delta(\rho_1 - \rho_2)\delta_{2\pi}(\phi_1 - \phi_2 - \pi),
\end{equation}
where $\rho$ and $\phi$ are the radial and azimuthal coordinates and $\delta_{2\pi}$ is the $2\pi$ periodic delta function. From 
this result we can expect the azimuthal angles 
of the photons in the far field to obey $\phi_1 = \phi_2 + \pi$, so that the signal and idler photons appear on opposite
sides of their respective cones. The correlation in angular momentum follows on writing Eq. (\ref{eq:anglecorr}) in
terms of its angular Fourier components:
\begin{equation}
\delta(x_1 + x_2)\delta(y_1 + y_2) = \frac{1}{\rho_1} \delta(\rho_1 - \rho_2) \frac{1}{2\pi} 
\sum_{m=-\infty}^\infty (-1)^m \exp(\rmi m \phi_1) \exp(-\rmi m \phi_2), 
\end{equation}
which is an entangled superposition of states with zero total OAM.
A more detailed analysis which considers a specific parametric down conversion process shows that more complicated dependences
of the wavefunction on the azimuthal angles are also possible \cite{barbosa+:pra65:2002}.

%%%%% SUBSECTION: EPR paradoxes 
\subsection{EPR paradoxes}
\label{ssc:eprpara}
The EPR paradox describes the apparent violation of the uncertainty principle resulting from measurements on
correlated, spatially separated systems. The original EPR argument considers correlations that are strong enough
to predict or infer with certainty the value of the observables in one subsystem from measurements on the 
other, separated, subsystem without disturbing in any way the first subsystem. The ability to predict with certainty the 
value of an observable defines, according to EPR, an element of reality. 
However, non-commuting observables, cannot have a simultaneous reality, an expression of which is the uncertainty principle 
\cite{robertson:pr34:1929}. The tension between local elements of reality and quantum complementarity leads to the paradox.

The original EPR \textit{Gedankenexperiment} considers an idealised situation. The quantum state given by 
EPR on the example of the position and momentum is -- in the modern language of entanglement -- a maximally entangled state 
\cite{nielsenchuang:CUP:2000}, and the measurement of the observables is assumed to be infinitely precise. For OAM and angular 
position this idealised setting would require 
a parametric down conversion process which creates an entangled photon pair, perfectly correlated in 
OAM and angular position. An errorless measurement of the OAM on the signal photon could then be used to infer the
OAM of the idler photon, and an errorless measurement of the azimuthal angle on the signal photon would allow
us to predict precisely the angle of the idler photon. As these measurements on the signal photon `do not
disturb the idler photon in any way', the predictions would constitute simultaneous elements of reality for the OAM 
and the azimuthal angle of the idler photon. We stress that the possibility to predict observables of the idler photon
with certainty depends on the ability to measure the observables on the signal photon without error.

In particular for a continuous observable a measurement with infinite precision cannot be realised experimentally. 
A typical experimental setup would allow us to determine whether a continuous variable falls into a previously specified range. 
To analyse the possibility of demonstrating an EPR paradox experimentally a more realistic situation has to be studied.
This requires the consideration of non-maximal correlations and of measurements with finite precision leading to an error 
in inferring one observable from a measurement on the other subsystem. The size of this error determines whether the EPR 
paradox can be demonstrated in the considered experimental setup \cite{reid:qso9:1997}.

%---------------------------

%%% SECTION: Formulation of the paradox

 \section{Formulation of the paradox}
The inclusion of experimental indeterminacies requires a reformulation of the EPR paradox. In contrast to the original EPR setup 
\cite{EPR:pr47:1935}, we consider conditional measurements on both subsystems. In reference to
the idealised setup in section (\ref{ssc:eprpara}), we look at the variance of the OAM or the azimuthal 
position in the idler beam given a specifically set outcome of a measurement in the signal beam. For the OAM this condition will 
be the measurement of a single $m$ value, whereas for the azimuthal position the measurement will be in a range of angles.
For the OAM we denote the conditional variance with $\var[m_2 | m_1]$, \ie the variance of $m_2$ in the idler under the condition
that a measurement on the signal photon yields an OAM of $m_1$. The condition for the azimuthal position to fall in a 
range of angles will be treated as a probability density $P_1(\phi_1;\tau_1)$. The functional dependence of the probability
density is given by $P_1(\phi_1)$ and the variable $\tau_1$ is used to indicate the orientation. For a symmetric
$P_1(\phi_1)$, $\tau_1$ would be the central angle.

The conditional variance for the angle can thus be
written as $\var[\phi_2 | P_1(\phi_1;\tau_1)]$.
For an actual experiment the error caused by non-maximum correlations is thus included in measured quantities and the
theoretical modelling, therefore, concentrates on the description of the error in the measurement. Schemes to measure
OAM have been theoretically studied and experimentally realised \cite{vaziri+:job4:2002,leach+:prl88:2002}. We are therefore 
mostly concerned with a 
description of the angle measurement. In the following we give the criterion for the experimental demonstration of an angular 
EPR paradox as one main result of this work.

%%% SUBSECTION: Criterion for  an angular EPR paradox
\subsection{Criterion for an angular EPR paradox}
\label{ssc:criterion}
In the previous section (\ref{ssc:eprpara}) we pointed out the importance of inference in EPR type arguments. To distinguish
a measured quantity from an inferred one we label the first with an index $\mathsf{m}$ and the latter with an index $\mathsf{i}$. The measured 
quantities are the conditional probability for the OAM $P[ m_2 | m_1]_{\mathsf{m}}$ and the conditional probability density for the azimuthal 
angle $P[ \phi_2 | P_1(\phi_1;\tau_1) ]_{\mathsf{m}}$.   
The paradox becomes now apparent if one assumes local realism. From the measured variance $P[ \phi_2 | P_1(\phi_1;\tau_1) ]_{\mathsf{m}}$ a minimum 
variance $\min \var [m_2 | P_1(\phi_1;\tau_1)]_{\mathsf{i}}$ can be derived which is still in accordance with the uncertainty relation 
[cf. Eq. (\ref{eq:oamur})]. This quantity can be compared to the measured variance $\var [m_2 | m_1]_{\mathsf{m}}$ 
as the conditioning measurement on the first subsystem `does not have an instantaneous influence on 
the second subsystem'. An angular EPR paradox would then be demonstrated if 
\begin{equation}
	\label{eq:paracrit}	
  \langle \var[m_2 | m_1]_{\mathsf{m}} \rangle_{m_1} < \langle \min \var[m_2 | P_1(\phi_1;\tau_1)]_{\mathsf{i}} \rangle_{\tau_1}. 
\end{equation} 
Within the simplified reasoning in section \ref{ssc:oam} the correlations in OAM are uniform and the correlations
in angle isotropic. This cannot be assumed a priori for an experimental test of the criterion Eq. (\ref{eq:paracrit}).
We are therefore considering averaged conditional variances, which take into account that the correlations may vary for
different values of $m_1$ and for different orientations of the angle $\tau_1$. 
In the quantity $\langle \var[m_2 | m_1]_{\mathsf{m}} \rangle_{m_1}$ the 
conditional OAM variance $\var[m_2 | m_1]_{\mathsf{m}}$ is averaged over the condition $m_1$. To find the average  
$\langle \min \var[m_2 | P_1(\phi_1;\tau_1)]_{\mathsf{i}} \rangle_{\tau_1}$ the minimum conditional variance $\min \var[m_2 | P_1(\phi_1;\tau_1)]_{\mathsf{i}}$
is integrated over $\tau_1$ on a $2\pi$ radian interval with the probability density $P(\phi_1;\tau_1)$. Formally the averaging
procedure eliminates the dependence of the criterion on particular values of $m_1$ and $\tau_1$, but not 
the dependence on the functional form of $P_1(\phi_1)$. The dependence on $P_1(\phi_1)$ originates from the way in which the condition
in Eq. (\ref{eq:paracrit}) is measured and is not directly connected to the correlation in the azimuthal angle.

With the usual definition of the variance in terms of probabilities, we can reformulate the paradox statement:
\begin{eqnarray}
    \sum_{m_1} |c_{m_1}|^2 & & \left[ \sum_{m_2} P[m_2 | m_1]_{\mathsf{m}} m_2^2 - \left( \sum_{m_2} P[m_2 | m_1]_{\mathsf{m}} m_2 \right)^2 \right] \nonumber \\
    & & < \int_{-\pi}^\pi d\tau_1 P_1(\phi_1;\tau_1) \min \left[ \sum_{m_2} P[m_2 | P_1(\phi_1;\tau_1)]_{\mathsf{i}} m_2^2 \right. \\
    & & \qquad \qquad - \left. \left( \sum_{m_2} P[m_2 | P_1(\phi_1;\tau_1)]_{\mathsf{i}} m_2 \right)^2 \right] \nonumber.
\end{eqnarray}
The inferred conditional probability $P[m_2 | P_1(\phi_1;\tau_1)]_{\mathsf{i}}$ will be calculated via a Fourier transform from 
$P[ \phi_2 | P_1(\phi_1;\tau_1) ]_{\mathsf{m}}$ and is given by the modulus square of the conditional
probability amplitudes $c[m_2 | P_1(\phi_1;\tau_1)]_{\mathsf{i}}$.  An experimental result which obeys the given
inequality [Eq. (\ref{eq:paracrit})] would constitute a demonstration of an angular EPR paradox.

%% % SUBSECTION: Angle measurement scheme
\subsection{Angle measurement scheme}
\label{ssc:anglemeas}
\begin{figure}
  \begin{center}
  \centerline{\epsfxsize=\textwidth\epsfbox{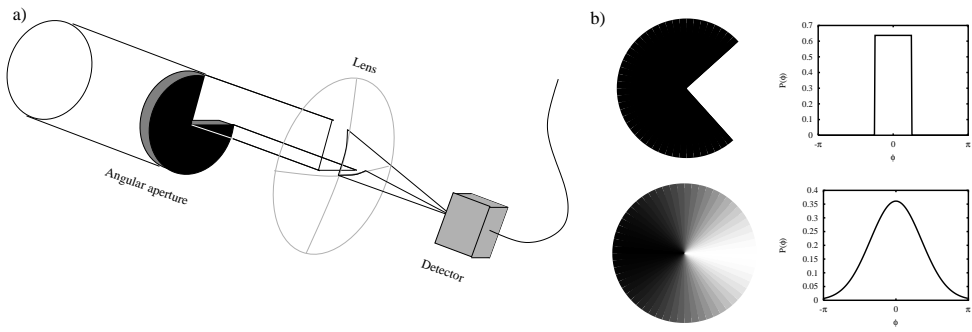}}
  \end{center}
  \caption{\label{fig:anglemeas}(a) Measurement scheme for the azimuthal position. A photon is said to have a 
  particular probability density for the azimuthal angle if it is detected after passing an aperture corresponding to 
  this probability density. Experimentally these apertures may be shaped using a spatial light modulator. (b) Aperture 
  functions and their associated azimuthal probability densities for a transmitted photon, shown for a rectangular 
  aperture and an aperture in form of a truncated Gaussian.} 
\end{figure}

The measurement scheme to determine the azimuthal position models experimental techniques employed in recent work
\cite{frankearnold+:njp6:2004}. A photon is said to have a specific angular probability density if it is detected 
after passing an angular aperture corresponding to this probability density [see Fig. \ref{fig:anglemeas}].
With the help of spatial light modulators smooth aperture functions can be realised.
 
A narrow aperture can be used to measure the probability density of the azimuthal position. The aperture could then be 
rotated, \ie the central angle could be varied over a $2\pi$ radian range. 
Detecting the number of photons passing the narrow aperture as a function of the central angle yields eventually a measure for
the probability distribution.

%%% SUBSECTION: Conditional variances
\subsection{Conditional variances}
\begin{figure}
  \begin{center}
  \epsfxsize=0.6\textwidth
  \epsfbox{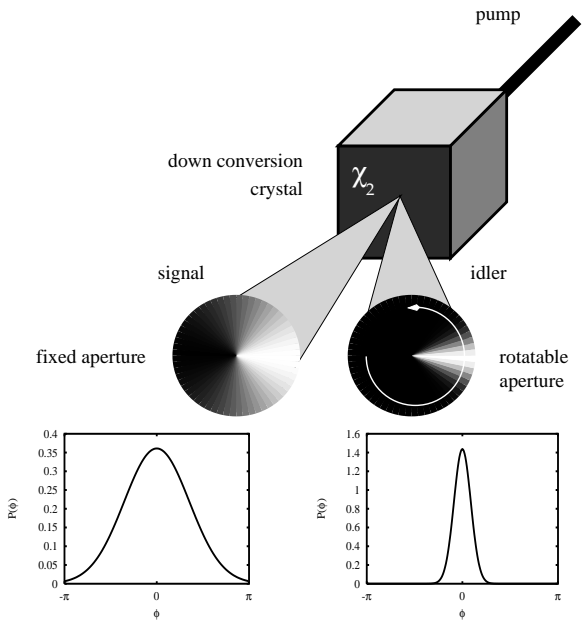}
  \end{center}
  \caption{\label{fig:condmeasure}Schematic representation of the conditional measurement. A fixed angular 
  aperture in the signal beam sets the condition and a rotatable aperture in the idler beam can be used to measure 
  a conditional probability density for the azimuthal position. The difference in the aperture functions  
  is chosen here for the purposes of illustration only.}
 \end{figure}

The angular apertures will also be used to set the condition in the signal beam. A fixed aperture can be inserted in 
the signal beam, and a rotatable aperture can be used to measure the conditional probability density in the idler beam 
[see Fig. \ref{fig:condmeasure}]. 

We will denote the outcome of such a measurement by $P[ \phi_2 | M_1(\tau_1) ]_{\mathsf{m}}$, \ie the probability density to detect an idler photon
with angle $\phi_2$ under the condition that the entangled photon in the signal beam passes the aperture $M_1$ oriented at $\tau_1$. 
As the aperture can be described by an aperture function, which we assume translates into a probability density $P_1(\phi_1; \tau_1)$, 
we will write synonymously $P[ \phi_2 | P_1(\phi_1; \tau_1) ]_{\mathsf{m}}$. The importance of the orientation angle $\tau_1$ has been stressed
in the formulation of the paradox [cf. section \ref{ssc:criterion}]. To simplify the notation we will not write the explicit
dependence of $P_1$ on $\tau_1$ from here on.

From the conditional probability density a class of 
wavefunctions in the angle representation $\psi[\phi_2 | P_1(\phi_1) ]$ is derived,
\begin{equation}
  \label{eq:condwave}
  \psi[\phi_2 | P_1(\phi_1) ] = \sqrt{ P[\phi_2 | P_1(\phi_1) ]_{\mathsf{m}}} \exp(\rmi \alpha(\phi_2)).
\end{equation}
Here, the phase $\rmi\alpha(\phi_2)$ is undetermined, as the measured probability densities only give the modulus square
of the wavefunction. The wavefunction is then transformed into a conditional OAM probability amplitude 
via a Fourier transform [cf. Eq. (\ref{eq:fouriertrans})]: 
\begin{equation} 
	c[m_2 | P_1(\phi_1)]_{\mathsf{i}} = \frac{1}{\sqrt{2\pi}} \int_{-\pi}^\pi d\phi_2 \exp(\rmi m \phi_2) 
	\sqrt{P[\phi_2 | P_1(\phi_1) ]_{\mathsf{m}}} \exp(\rmi \alpha(\phi_2)).
\end{equation}
From the conditional probability amplitudes we can calculate the conditional variance $\var [m_2 | P(\phi_1)]_{\mathsf{i}}$ by taking 
the sum over all $m_2$ values
\begin{equation}
	\label{eq:condvar}	
\min \var [m_2 | P(\phi_1)]_{\mathsf{i}} = \min \left[ \frac{\sum_{m_2} c[m_2 | P_1(\phi_1)]^2_{\mathsf{i}} l^2 }
{\sum_{m_2} c[m_2 | P_1(\phi_1)]^2_{\mathsf{i}}} - \left[ \frac{\sum_{m_2} c[m_2 | P_1(\phi_1)]^2_{\mathsf{i}} l}
{\sum_{m_2} c[m_2 | P_1(\phi_1)]^2_{\mathsf{i}}} \right]^2 \right].
\end{equation}
This is the inferred minimum variance which can be compared to the measured quantity $\var[ m_2 | m_1]_{\mathsf{m}}$. The phase 
$\rmi\alpha(\phi_2)$ will be determined by the minimization of the conditional variance $\var [m_2 | P_1(\phi_1)]_{\mathsf{i}}$ as
detailed in the following section. 

%%%%% SUBSECTION: Minimization of the conditional variance
\subsection{Minimization of the conditional variance}
Calculating the conditional wavefunction from the conditional probability leaves the phase $\rmi\alpha(\phi_2)$ undetermined. 
We find that if the variance is calculated for $\alpha(\phi_2) \equiv 0$ then the minimum variance $\min \var [m_2 | P_1(\phi_1)]_{\mathsf{i}}$ is 
obtained. 
To show this we assume the conditional wavefunction to be of the form
\begin{equation}
  \psi [\phi_2 | P_1(\phi_1) ] = A \exp[\rmi \alpha(\phi_2)],
\end{equation}
where $A=\sqrt[+]{|\psi[\phi_2 | P_1(\phi_1)]|^2}$ is a positive, real function, which is periodic in $\phi_2$. Applying the orbital 
angular momentum operator to the wavefunction yields
\begin{eqnarray}
  L_z \psi[\phi_2 | P_1(\phi_1)] & = & -\rmi \hbar \psi'[\phi_2 | P_1(\phi_1)] \nonumber \\
& = & -\rmi \hbar A'(\phi) \exp[ \rmi \alpha(\phi_2) ] + \hbar \alpha'(\phi_2) \exp[\rmi \alpha(\phi_2)],
\end{eqnarray}
where the primes denote derivatives with respect to $\phi_2$. Using the periodicity of $A$ the variance of $L$ can be evaluated to
\begin{eqnarray}
  \var L_z &  = &  \hbar^2 \int d\phi_2 A(\phi_2) A''(\phi_2)  + \hbar^2 \int d\phi_2 (\alpha'(\phi_2))^2 P(\phi_2) -  \nonumber \\
  & = & \hbar^2 \left( \int d\phi_2 \alpha'(\phi_2) P(\phi_2) \right)^2,
\end{eqnarray}
where we used the fact that $A^2(\phi_2) = P[\phi_2 | P_1(\phi_1)]_{\mathsf{m}}$. The first integral is the variance of $L$ for $\alpha(\phi_2) \equiv 0$, while the 
second and third integral are the variance of $\alpha'$:
\begin{equation}
\var L_z = [\var L_{z}]_{\alpha = 0} + \hbar^2 \var \alpha'.
\end{equation}
Therefore we obtain the minimum variance if the wavefunction $\psi[\phi_2 | P(\phi_1)]$ is real and positive, so that $\alpha = 0$. 
In the following we will consider only this case.

%----------------------------

 %%% SECTION:  Proposed experimental scheme
\section{Proposed experimental scheme}
\begin{figure}
\begin{center}
\centerline{\epsfxsize=0.8\textwidth\epsfbox{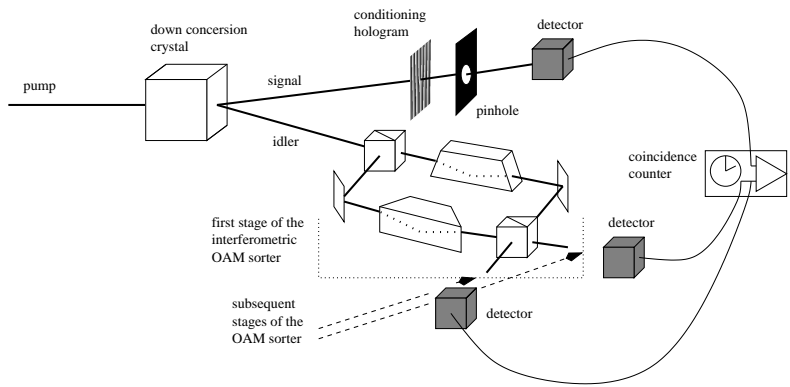}}
\caption{\label{fig:condexp1} Experimental scheme for measuring $\var[m_2 | m_1]_{\mathsf{m}}$. In the signal a hologram is used to single out a particular 
value of $m_1$ as a condition. The OAM distribution of the idler is determined with help of an interferometric OAM sorter
\cite{leach+:prl88:2002}. Only the first stage of the sorter is shown here, additional stages are added where indicated by
arrows. Each stage doubles the possible outcomes and therefore the number of detectors. Eventually the signals from all detectors
are transmitted to a coincidence counter.} 
\end{center}
\end{figure}

In order to measure $\var[ m_2 | m_1 ]_{\mathsf{m}}$ we have to examine the signal photon for the particular OAM $m_1$ and determine
the OAM $m_2$ of the idler photon. A schematic representation of an experimental setup to achieve this is shown in 
Fig. (\ref{fig:condexp1}). A spatial light modulator is used to produce a hologram which 
changes the OAM in the signal photon by the chosen value $-m_1$ to zero 
\cite{vaziri+:job4:2002,frankearnold+:njp6:2004}. Only beams with $m_1=0$ have on-axis intensity and can thus
be detected behind a pinhole. The measurement for the idler photon has to distinguish between different values
of $m_2$. A sorting scheme which is able to determine the OAM has been experimentally implemented 
\cite{leach+:prl88:2002}. A coincidence measurement would then yield $\var[m_2 | m_1]_{\mathsf{m}}$.

Experimentally it will not be possible to measure the conditional probability density for a single angle, instead a suitable aperture 
can be used to test for a range of angles, as described in section \ref{ssc:anglemeas}. Analogously to the conditioning aperture, the 
measured quantity is $P [ M_2(\tau_2) | P_1(\phi_1) ]_{\mathsf{m}}$, where the aperture $M_2$ is centred at a particular angle $\phi_2 = \tau_2$.
The aperture can be described by a probability density $P_2(\phi_2;\tau_2)$, where $\tau_2$ indicates the central angle.
The measured probability can thus be written as $P[ P_2(\phi_2;\tau_2) | P_1(\phi_1)]_{\mathsf{m}} $.
For a very narrow aperture $M_2$ this measurement will give a good estimate of $P [ \tau_2 | P_1(\phi_1)]_{\mathsf{m}}$
\begin{equation}
P [ \tau_2 | P_1(\phi_1)]_{\mathsf{m}} \approx P [ P_2(\phi_2;\tau_2) | P_1(\phi_1)]_{\mathsf{m}} \quad \mathrm{for}\:\,\mathrm{suitable}\:\, P_2(\phi_2).
\end{equation}
The error made in this approximation is then given by the variance of $\phi_2$ for the probability density $P_2$.
   	
The scheme to measure $P[P_2(\phi_2;\tau_2) | P_1(\phi_1)]_{\mathsf{m}}$ is basically shown in Fig. (\ref{fig:condmeasure}). With the help of a
spatial light modulator [SLM] any chosen $P_1(\phi_1)$ can be set as a condition for the signal. 
For the idler the aim is to determine $\phi_2$ as exactly as possible. To achieve this an SLM could be programmed for
a narrow angular aperture $P_2(\phi_2;\tau_2)$ centred at $\phi_2 = \tau_2$, which would then be varied over the $2\pi$ radian interval. 
Eventually this would lead to $P[P_2(\phi_2) | P_1(\phi_1)]_{\mathsf{m}}$ which does not only depend on the condition $P_1(\phi_1)$ but also
on the chosen analysing aperture $P_2(\phi_2)$. This is the experimentally measurable estimate of the quantity $P[\phi_2 | P_1(\phi_1)]_{\mathsf{m}}$ 
used
in the formulation of the angular EPR paradox [cf. section (\ref{ssc:criterion})]. Obviously there are experimental 
limitations: a narrow aperture would transmit only little intensity, which would reduce the detection rate in the conditional 
measurement. Also SLMs have a finite size and resolution, which limits the ability to distinguish between similar apertures.

The interpretation of experimental data would require the inclusion of imperfect correlations originating from the parametric
down conversion process and the influence of the optical elements, in particular the angular apertures. In the following
we will discuss these aspects briefly.

%%%%% SUBSECTION: Parametric down conversion
\subsection{Parametric down  conversion}
The possibility to create photon pairs entangled in OAM or angular position is based on the conservation of 
OAM under parametric down conversion \cite{mair+:nat412:2001,frankearnold+:pra65:2002}, which holds for thin down conversion
crystals and in the paraxial limit. The conservation leads to a perfect anti-correlation in the OAM indices so that for
a given OAM index $m_p$ in the pump, signal and idler obey $m_p = m_1 + m_2$. In a recently reported down conversion experiment
\cite{altman+:qph:2004}, the spatial correlations of photons entangled in orbital angular momentum have been studied. In this
particular experiment signal and idler cone overlap completely and the spatial correlations are such that for a fixed detector 
position in the signal, the coincidence pattern in the idler shows two distinct spots equally separated in angle from the position 
exactly opposite the signal detector on the phase matching ring. The vertex of the separation angle is on the pump axis. 
In our work we are concerned with a different angle: the azimuthal position of a photon in a beam is measured from
the beam axis, \ie in a down conversion experiment from the signal and idler axis respectively. For non degenerate down conversion
crystals these two angles are not identical. The question if they are compatible for the degenerate case is a very interesting one,
but it will not be examined in the scope of this work. Therefore, for our theoretical modelling, we make use of the simplified 
reasoning given in section (\ref{ssc:oam}) and therefore have an angle correlation $\phi_1 - \phi_2 = \pi$ in the far field
[cf. Eq. (\ref{eq:anglecorr})]. 
 
A recent study on the EPR paradox for linear optical momentum and position \cite{howell+:prl92:2004} included a theoretical 
prediction of the conditional probabilities. Although this is certainly an interesting additional information for interpreting 
experimental data, the analysis here does not rely on it, as we would like to end with a criterion that can be applied directly to 
experimental data. The effect of imperfect correlations is therefore included in the measured quantities $\var[m_2 | m_1]_{\mathsf{m}}$ 
and $P[P_2(\phi_2) | P_1(\phi_1) ]_{\mathsf{m}}$.  

%%%%% SUBSECTION: Angular aperture using spatial light modulators
\subsection{Angular apertures using spatial light modulators}
\label{ssc:aperslm}
The advantage of spatial light modulators [SLMs] is that, within the spatial resolution of the SLMs, angular apertures may be 
smooth functions of the angle $\phi$. Rectangular functions, which represent `cake-slice' apertures, would result in singular
derivatives and hence in an 
infinite inferred variance $\var[m_2 | P_1(\phi_1)]_{\mathsf{i}}$ \cite{pegg+:njp7:2005}. However, this analysis does not take into account 
that optical diffraction will have a smoothing effect on the angle probability distribution. To study the influence of even very 
small smoothing effects we consider in the following theoretical modelling a class of continuously differentiable aperture functions 
which asymptotically approximate rectangular functions. On the other hand, aperture functions which differ only slightly may 
be mapped to the same aperture in the SLM, because of the limited resolution.
These aspects have to be discussed more closely in conjunction with specific experimental implementations. 

%----------------------------

%%% SECTION: Theoretical modelling
\section{Theoretical modelling}
To give a quantitative result we model the measurement of $P[P_2(\phi_2) | P_1(\phi_1)]$ under the two assumptions that
the photon pair is perfectly correlated in angular position and that the angle probability distribution
behind the aperture is exactly given by the function describing the aperture. Under these assumptions the conditional
probability $P[P_2(\phi_2) | P_1(\phi_1)]$ is given by the overlap integral of the two probability densities, since the detection probability
for an analyzing aperture centred at $\phi_2 = \tau_2$ is given by
\begin{equation}
P_{\mathrm{dtc}}(\phi_2 = \tau_2) = \int_{-\pi}^{\pi}\int_{-\pi}^{\pi} |\psi(\phi_1, \phi_2)|^2 P_1(\phi_1) P_2(\phi_2; \tau_2) 
d \phi_1 d \phi_2 .
\end{equation}
Using the assumption of perfect correlation the probability density will be sharply peaked for $\phi_1 - \phi_2 = \pi$ and
in that sense we may use a $2\pi$-periodic $\delta$-function to approximate $|\psi(\phi_1, \phi_2)|^2$
\begin{equation}
|\psi(\phi_1, \phi_2)|^2 \approx \delta_{2\pi} (\phi_1 - \phi_2 - \pi).
\end{equation}
The idler probability density $P_2(\phi_2; \tau_2)$, may be written as a function centred at $\phi_2 = 0$ but shifted by
$\tau_2$, which yields $P_2(\phi_2 - \tau_2)$. Using these results the detection probability can be rewritten as
\begin{equation}
\label{eq:overlap}	
  P_{\mathrm{dtc}}(\phi_2 = \tau_2) = \int_{-\pi}^{\pi} P_1(\phi_1) P_2(\phi_1 + \pi - \tau_2) d\phi_1.
\end{equation}
For the calculation of this integral it is important to use the periodicity of the probability densities if the argument
lies outside the $2\pi$ radian interval.  The detection probability $P_{\mathrm{dtc}}(\phi_2 = \tau_2)$ is the measured
conditional probability $P[P_2(\phi_2;\tau_2) | P_1(\phi_1)]_{\mathsf{m}}$. Varying $\tau_2$ over the $2\pi$ radian interval yields
the conditional probability $P[P_2(\phi_2) | P_1(\phi_1)]_{\mathsf{m}}$. In an experimental test of our criterion [Eq. \ref{eq:paracrit}] this 
quantity would be known from measurements.

%%%%% SUBSECTION: Aperture functions
\subsection{Aperture functions}
We model the conditional angle measurement for different aperture functions. As experiments to validate the
uncertainty relation \cite{frankearnold+:njp6:2004} used apertures which can be described by truncated Gaussians, we
calculate the effects of these truncated Gaussian apertures and also of a set of truncated super Gaussians. The latter
allows us to interpolate between Gaussian and rectangular apertures. As mentioned in section (\ref{ssc:aperslm}) rectangular apertures, 
which could be made from a solid, absorbent material, will lead to an infinite inferred variance \cite{pegg+:njp7:2005}.  
All aperture functions discussed in this section are symmetric, \ie $P_j(\phi_j) = P_j(-\phi_j)$ for $j=1,2$. The overlap integral in
Eq. (\ref{eq:overlap} )can therefore be turned into the convolution of the two probability densities
\begin{equation}
\label{eq:convolution}
P_{\mathrm{dtc}}(\phi_2 = \tau_2) = \int_{-\pi}^{\pi} P_1(-\phi_1) P_2(\phi_1 + \pi - \tau_2) d\phi_1 = [P_1 \ast P_2](\pi - \tau_2).
\end{equation}
For simplicity we are modelling the quantity $P[ \phi_2 | P_1(\phi_1)]_{\mathsf{m}}$ with the convolution 
$[P_1 \ast P_2](\phi_2)$ as for the symmetric and 
periodic aperture functions the shift by $\pi$ and the sign of $\tau_2$ is not relevant. 

%%%%%%%%% SUBSUBSECTION: Rectangular functions
\subsubsection{Rectangular functions}
\begin{figure}
  \begin{center}
    \epsfxsize=0.49\textwidth
    \epsfbox{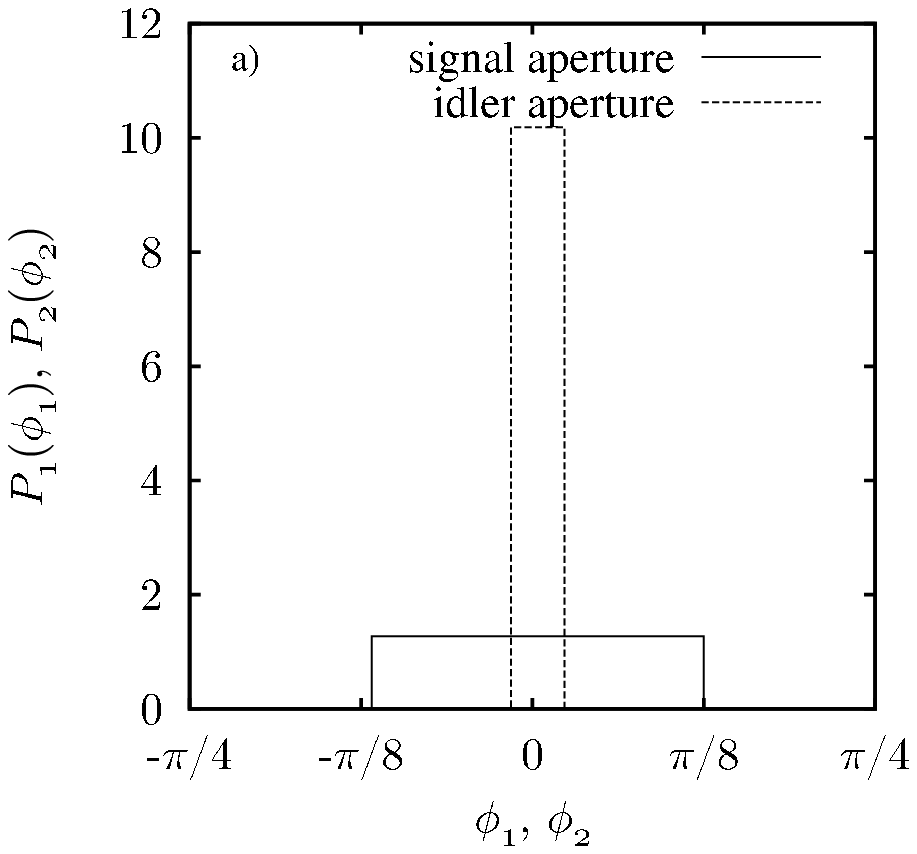}
    \epsfxsize=0.49\textwidth
    \epsfbox{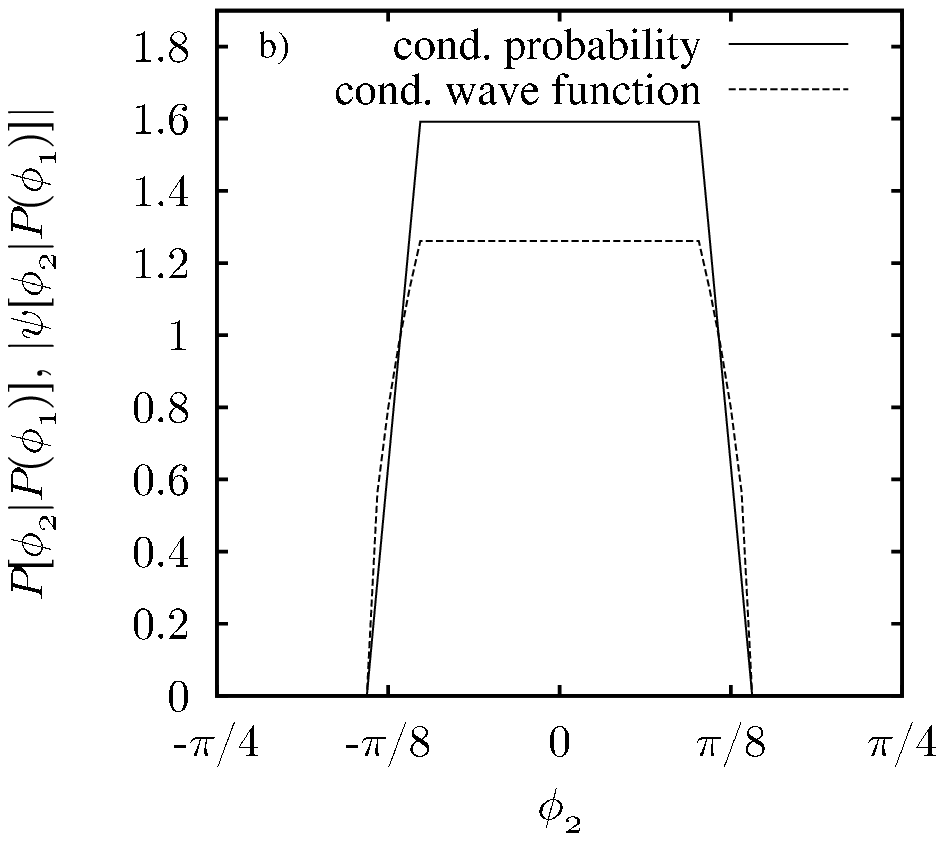}
  \end{center}
  \caption{\label{fig:rect12} In $a)$ the probability densities representing the apertures in the signal and idler beam are plotted,
	for a width in the idler and signal of $w_1 = 1/4 \pi$ and $w_2 = 1/64 \pi$ respectively. The resulting conditional probability 
	density and wavefunction are shown in $b)$. The wavefunction is taken as the modulus of the probability density as a purely real 
        wavefunction minimizes the OAM variance.}
\end{figure}

The probability functions describing rectangular apertures can be given in terms of Heaviside step functions $H(\phi)$:
\begin{equation}
  P_j(\phi_j) = \frac{1}{w_j} H(\phi_j+\frac{w_j}{2})H(-\phi_j+\frac{w_j}{2}) \quad \mathrm{for}\: j=1,2.
\end{equation}  
According to Eq. (\ref{eq:convolution}) the conditional probability is given by the convolution of the two
probability densities
\begin{equation}
	[P_1 \ast P_2](\phi_2) = \frac{1}{\Delta_1^2 - \Delta_2^2} 
	\left\{ \begin{array}{ll} 
		\Delta_1 - \Delta_2 	& |\phi_2| < \Delta_2, \\
		\Delta_1 - \phi_2	& \Delta_2 \leq |\phi_2| < \Delta_1, \\
		0			& \Delta_1 \leq |\phi_2| 
	\end{array} \right.
\end{equation}
where we introduced the notation $\Delta_1 = (w_1 + w_2)/2$ and $\Delta_2 = (w_1 - w_2)/2$ for $w_1 \geq w_2$. As the
convolution is commutative [$P_1 \ast P_2 = P_2 \ast P_1$] we can interchange $w_1$ and $w_2$ for the case $w_2 < w_1$. 
The conditional wavefunction is then given by the positive square root. Graphs of the rectangular apertures and the resulting conditional probability 
density and wavefunction are shown in Fig (\ref{fig:rect12}).

\begin{figure}
  \begin{center}
    \epsfxsize=0.49\textwidth
    \epsfbox{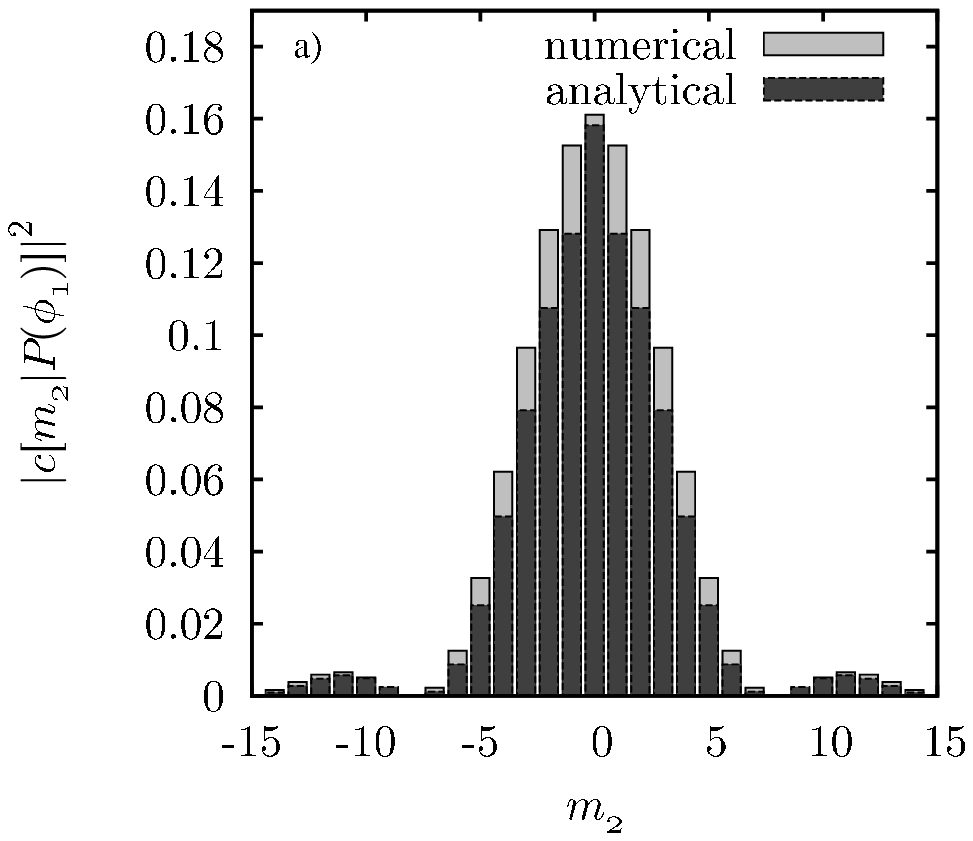}
    \epsfxsize=0.49\textwidth
    \epsfbox{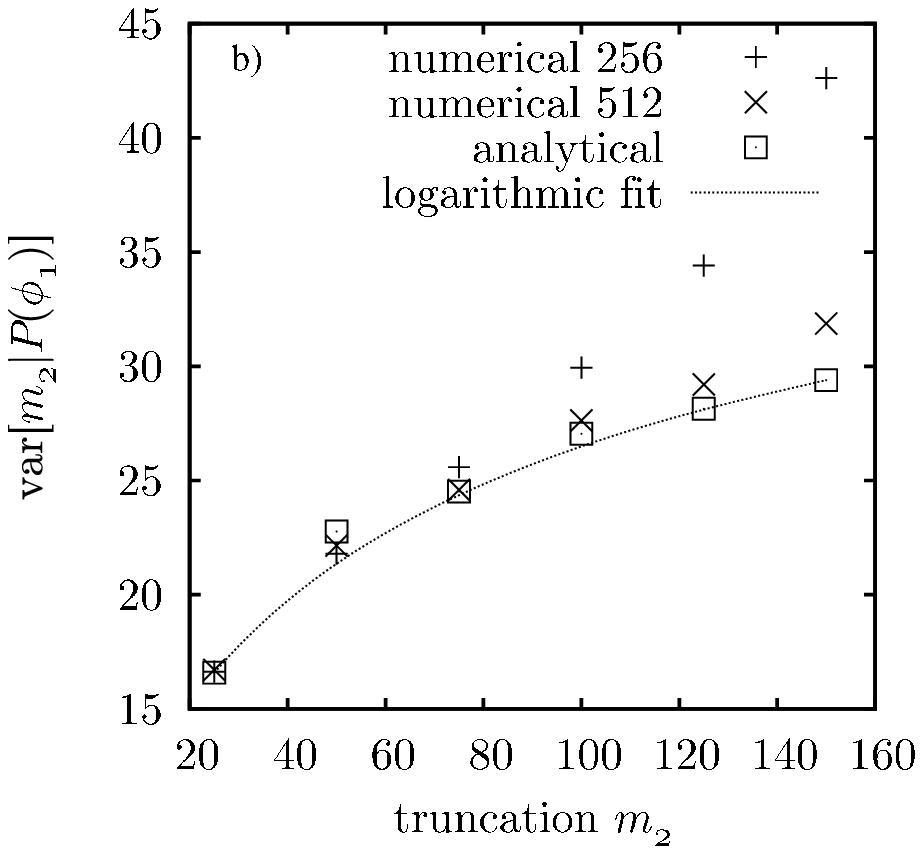}
  \end{center}
  \caption{\label{fig:rect34}$a)$ Numerically and analytically calculated OAM distribution for rectangular apertures. The analytical
  solution is exact, but to give an estimate of the accuracy of our integration method, the numerical results are also shown. 
  In $b)$ the conditional variance is plotted. The numerical results are given for the two different sampling sizes 256 and 512.
  The deviation of the numerical 256 results for higher truncation indices is caused by numerical effects. For the analytical points the
  logarithmic increase in the variance can be seen.}
\end{figure}

The Fourier integral [Eq. (\ref{eq:fouriertrans})] can be calculated analytically by using the Fresnel sine and cosine integrals $\mathcal{S}_2$ and
$\mathcal{C}_2$ \cite{gradshteyn+:ap:2000}:
\begin{equation}
\begin{array}{rl}
  c[ m_2 | P_1(\phi_1)] = \frac{1}{\sqrt{\Delta_1 - \Delta_2}}
  \times \bigg[ & \frac{\sin(|m_2| \Delta_1)}{|m_2| \sqrt{|m_2|}} \mathcal{C}_2( |m_2|(\Delta_1-\Delta_2)) \\
  & -\frac{\cos(|m_2| \Delta_1)}{|m_2| \sqrt{|m_2|}} \mathcal{S}_2( |m_2|(\Delta_1-\Delta_2)) \bigg].
\end{array}
\end{equation}
The distribution of conditional probability amplitudes is given in Fig. (\ref{fig:rect34}.a) for the analytical
calculation and a numerical integration. The numerical results are shown to give an estimate of the 
accuracy of our integration. This serves as a reference for aperture functions, where we have not found an analytical 
solution of the 
Fourier integral. The conditional variance is shown in Fig. (\ref{fig:rect34}.b) over a maximum OAM index $m_2$
at which the sum in Eq. (\ref{eq:condvar}) is truncated. The numerical results differ from the analytical points
at higher truncation indices because of numerical effects in sampling the $2\pi$ radian interval. Since the Fresnel integrals
tend to $1/2$ for large arguments, the conditional amplitudes $c [m_2 | P_1(\phi_1)]$ vary with $|m_2|^{-3}$ for large
$m_2$, which leads to a logarithmic increase with higher truncation indices and thus to a divergent conditional
variance. The reason for this behaviour is founded in the singular derivative of the conditional wavefunction
at $\phi_2 = \Delta_1$.

%%%%%%%%% SUBSUBSECTION: Truncated Gaussians 
\subsubsection{Truncated Gaussians}
\begin{figure}
  \begin{center}
    \epsfxsize=0.49\textwidth
    \epsfbox{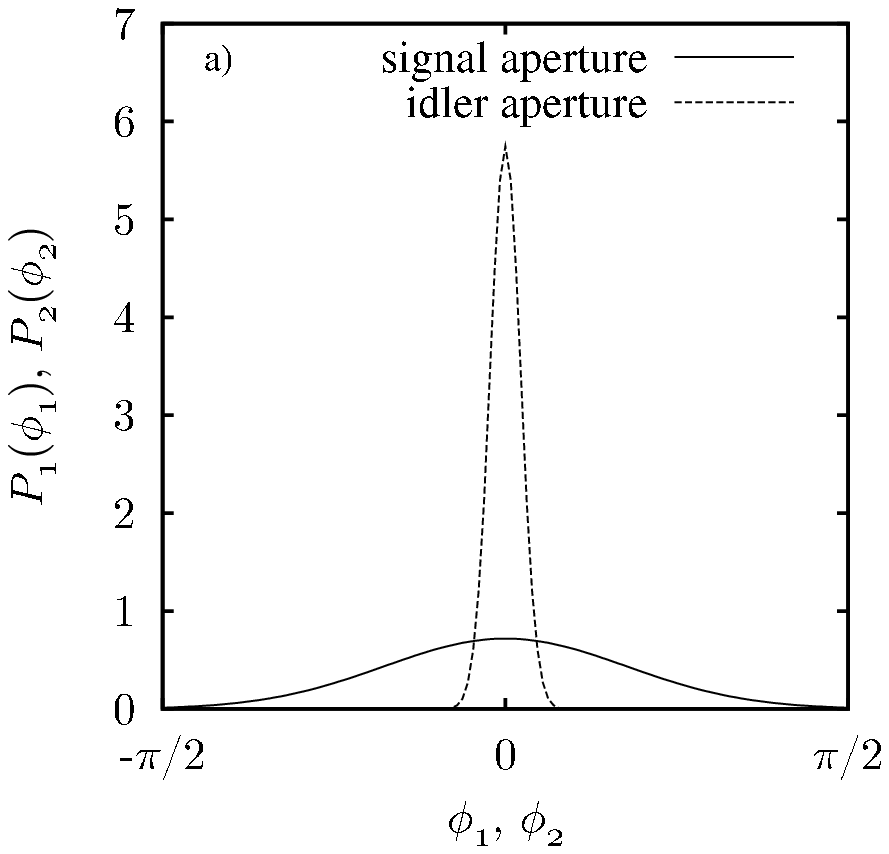}
    \epsfxsize=0.49\textwidth
    \epsfbox{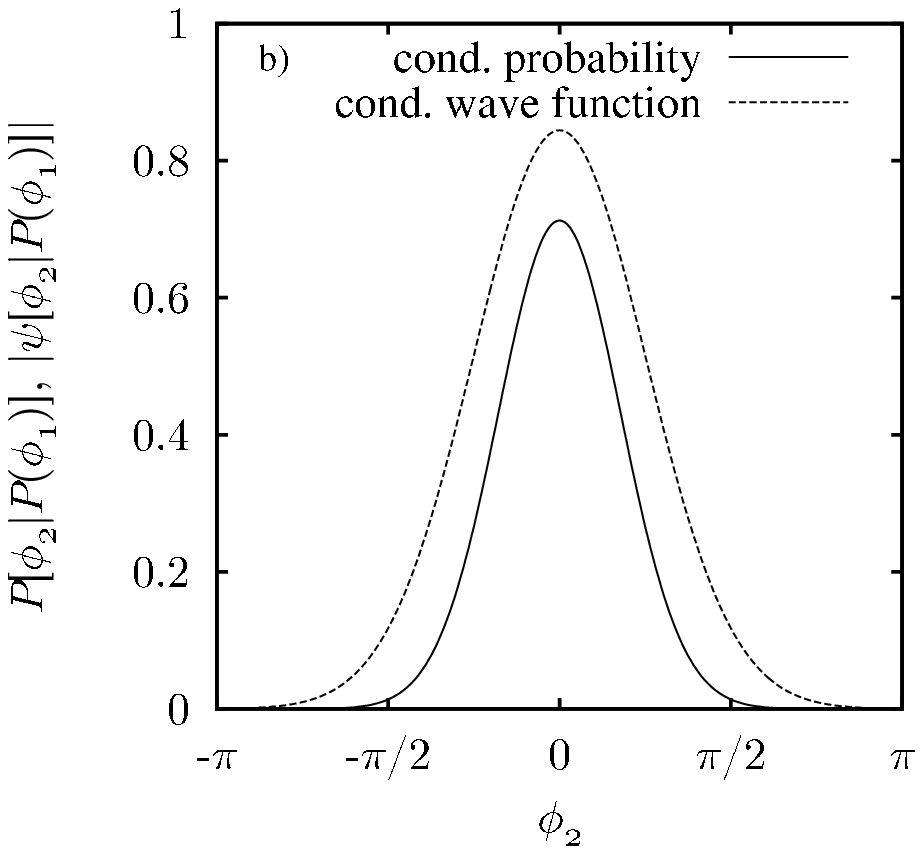}
  \end{center}
  \caption{\label{fig:gaus12} In $a)$ the probability densities representing the apertures in the signal and idler beam are plotted,
	for a width in the idler and signal of $w_1 = 1/4 \pi$ and $w_2 = 1/64 \pi$ respectively. The resulting conditional probability
	density and wavefunction are shown in $b)$. The wavefunction is taken as the modulus of the probability density as a purely real 
        wavefunction minimizes the OAM variance.}
\end{figure}

Aperture functions in form of truncated Gaussians have been used in the experiment studying the angular uncertainty
principle \cite{frankearnold+:njp6:2004}. In this case the probability densities $P_1$ and $P_2$ are
\begin{equation}
	P_j(\phi_j) = \frac{1}{\sqrt{\pi}w_j\erf(\pi/w_j)} \exp \left( -\left(\frac{\phi_j}{w_j}\right)^2 \right) \quad\mathrm{for}\: 
	\phi_j \in [-\pi,\pi), j=1,2.
\end{equation}
The graph for Gaussian apertures is shown in Fig. (\ref{fig:gaus12}.a) for the same widths $w_1$ and $w_2$ as
in the rectangular case. The resulting conditional probability density and wavefunction are plotted in 
Fig. (\ref{fig:gaus12}.b).

\begin{figure}
  \begin{center}
    \epsfxsize=0.49\textwidth
    \epsfbox{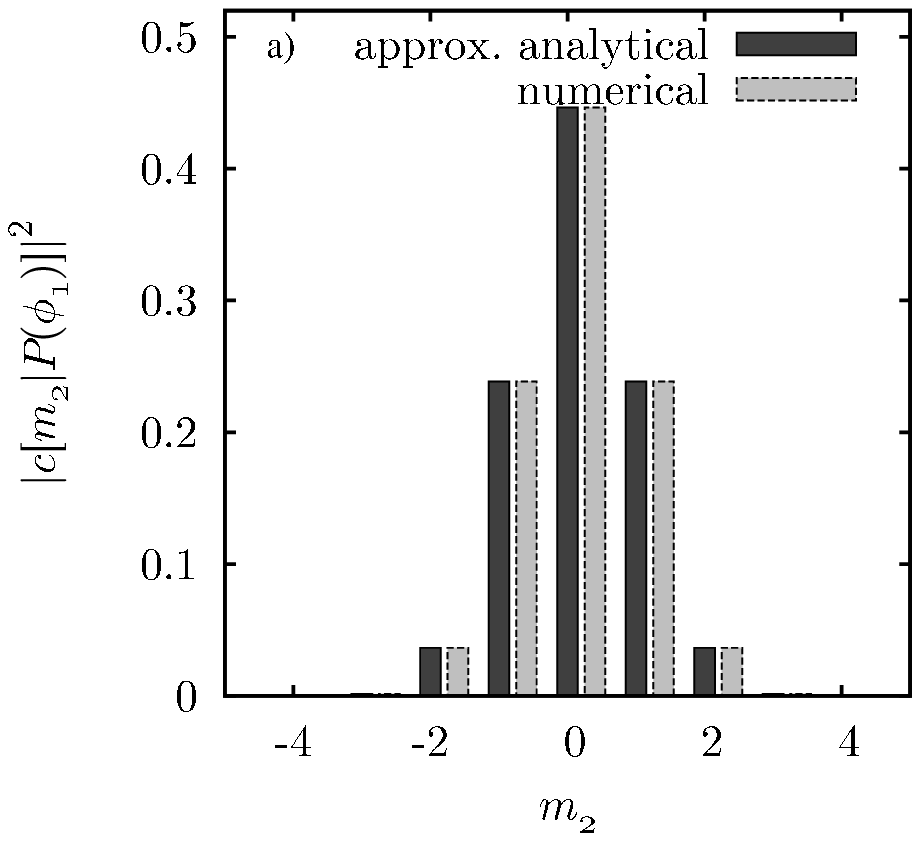}
    \epsfxsize=0.49\textwidth
    \epsfbox{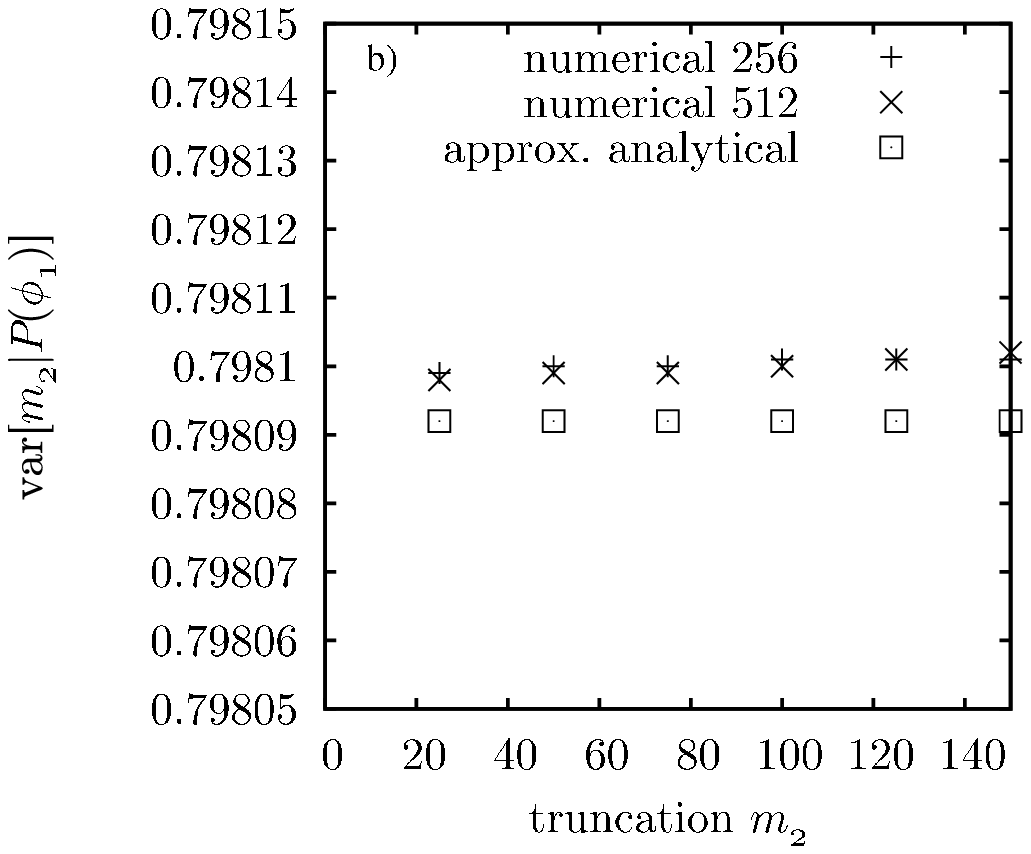}
  \end{center}
  \caption{\label{fig:gaus34}$a)$ Numerically and analytically calculated OAM distribution for truncated Gaussians apertures. 
  The analytical
  solution is approximated by neglecting the finite interval and treating the truncated Gaussians as infinite. This allows us to
  calculate the probability amplitudes using the convolution theorem and the fact that square roots of Gaussians are again
  Gaussians. The resulting conditional variance can be seen in $b)$. For the Gaussian apertures we have a constant conditional
  variance.} 
\end{figure}

The conditional probability density $P[\phi_2 | P(\phi_1)]$ can be analytically calculated using the convolution
theorem for the Fourier transform
\begin{eqnarray}
\label{eq:gaussconvolution}
		\frac{1}{\sqrt{2\pi}} \int_{-\pi}^\pi d\phi_2 & &\exp(\rmi m_2 \phi_2) [P_1 \ast P_2](\phi_2) = \frac{1}{\sqrt{2\pi}} 
		\frac{1}{\erf(\pi/w_1)\erf(\pi/w_2)} \nonumber \\
		& & \times \exp \left( -\frac{m_2^2(w_1^2 + w_2^2)}{4} \right) \\
	 & & \times \Real\left[\erf\left( \frac{\pi}{w_1} - \rmi \frac{m_2 w_1}{2}\right)\right]
	\Real\left[\erf\left( \frac{\pi}{w_2} - \rmi \frac{m_2 w_2}{2}\right)\right]. \nonumber
\end{eqnarray}
An approximate analytical expression for the probability amplitudes can be obtained by treating the Gaussians
as extended which results in setting the error functions to unity. This allows us to
calculate the $c[m_2 | P_1(\phi_1)]$ from Eq. (\ref{eq:gaussconvolution})
\begin{equation}
	c[ m_2 | P_1(\phi_1)] \approx \left( \frac{w_1^2 + w_2^2}{\pi} \right)^{\frac{1}{4}}
	\left( \frac{1}{\erf(\pi/w_1)\erf(\pi/w_2)} \right)^{\frac{1}{2}}
	\exp\left( - \frac{m_2^2(w_1^2 + w_2^2)}{2} \right).
\end{equation}
For the chosen widths the agreement with the numerical solution is excellent as can be seen in 
Fig. (\ref{fig:gaus34}.a). This is, however, a singular case and for different parameter settings the 
approximation differs more from the numerical solution.

The resulting conditional variance is shown for numerical and analytical calculations. For the truncated Gaussians the 
variance converges. The difference between the numerical results and the analytical solution is relatively small, because
the contributions from larger values of $m_2$ are very small.

%%%%%%%%% SUBSUBSECTION: Truncated super Gaussians 
\subsubsection{Truncated super Gaussians}

\begin{figure}
  \begin{center}
    \epsfxsize=0.49\textwidth
    \epsfbox{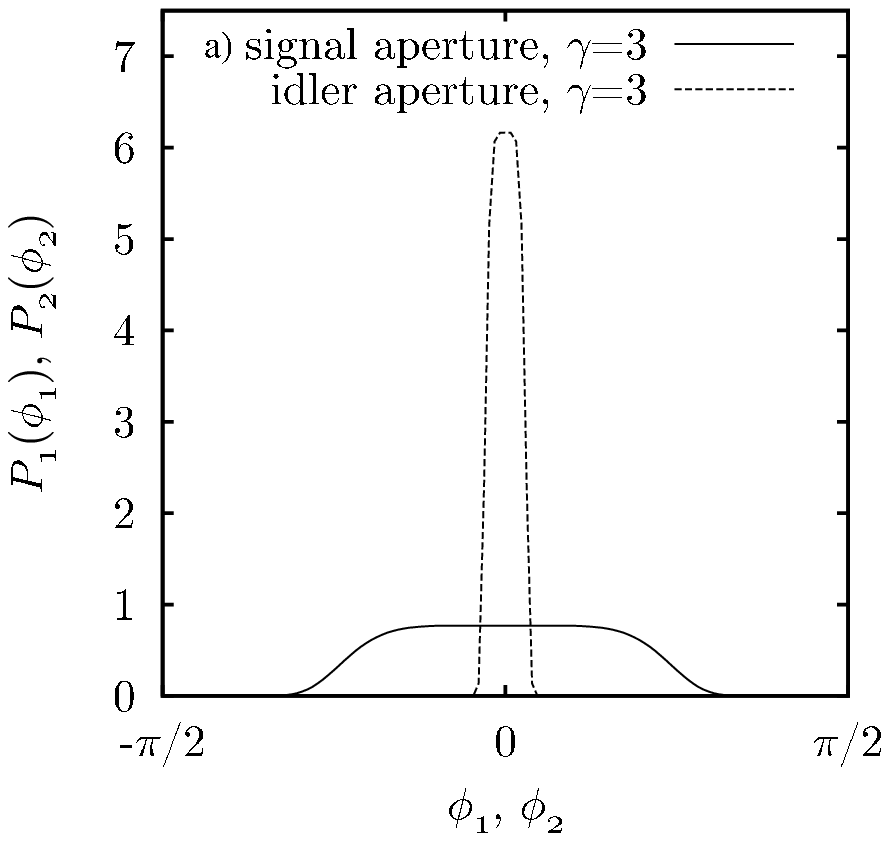}
    \epsfxsize=0.49\textwidth
    \epsfbox{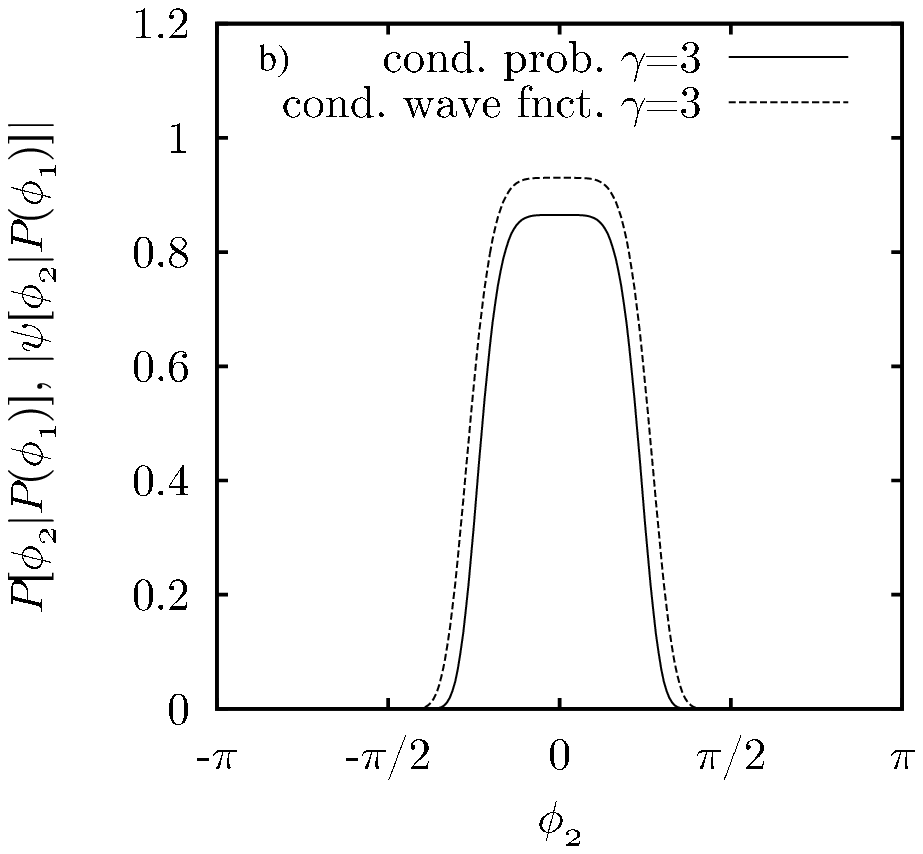}
    \epsfxsize=0.49\textwidth
    \epsfbox{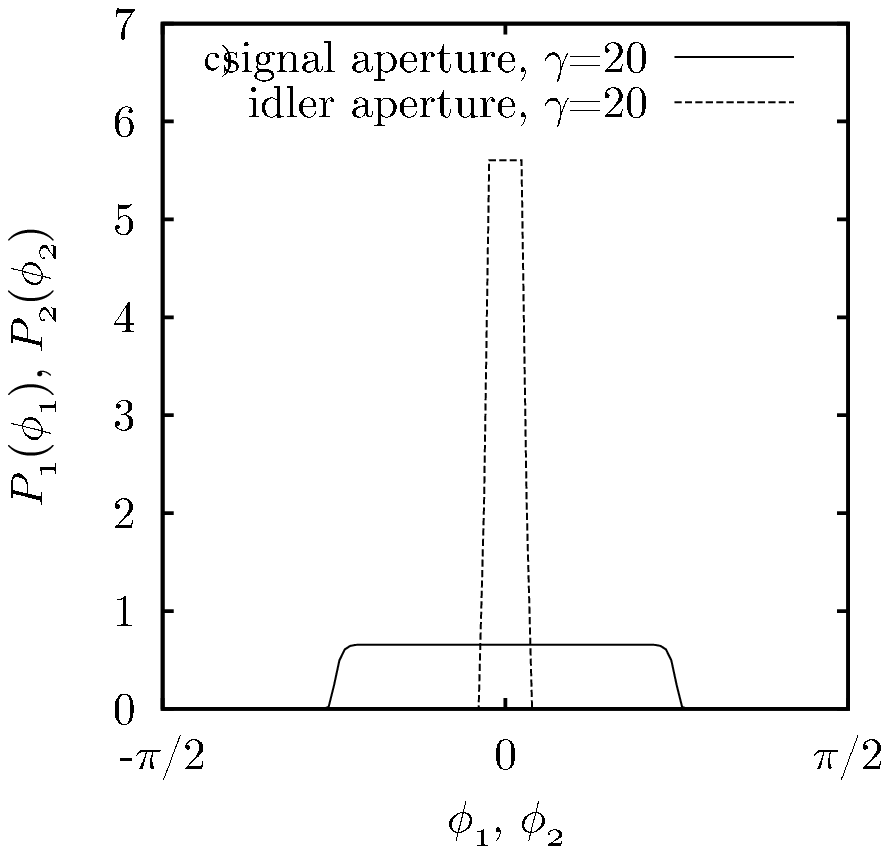}
    \epsfxsize=0.49\textwidth
    \epsfbox{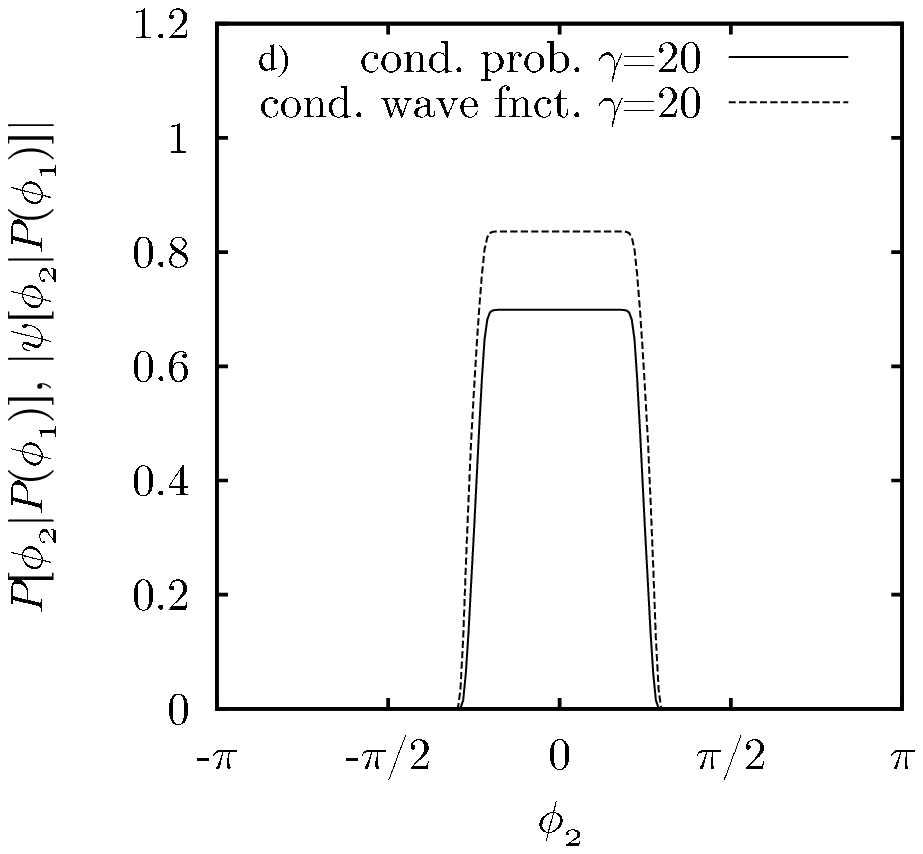}
  \end{center}
  \caption{\label{fig:suga12} $a)$ Probability densities representing the apertures in the signal and idler beam 
	[$w_1 = 1/4 \pi$, $w_2 = 1/64 \pi$, $\gamma=3$]. $b)$ Resulting conditional probability density and wavefunction 
        for $\gamma=3$. $c)$ Probability densities representing the apertures in the signal and idler beam 
	[$w_1 = 1/4 \pi$, $w_2 = 1/64 \pi$, $\gamma=20$]. $d)$ Resulting conditional probability density and wavefunction 
        for $\gamma=20$}.
\end{figure}

With truncated super Gaussian [TSG] apertures we can gradually go from the rectangular case to the Gaussian. This is achieved by an
additional parameter $\gamma$ in our definition of the TSG:
\begin{equation}
	P_j(\phi_j) = \frac{\gamma}{w_j( \Gamma(\frac{1}{2}\gamma) - \Gamma(\frac{1}{2}\gamma; (\frac{\pi}{w_j})^{2\gamma})} 
	\exp \left( -\frac{\phi_j^{2\gamma}}{w_j^{2\gamma}} \right) \quad\mathrm{for}\: 
	\phi_j \in [-\pi,\pi), j=1,2,
\end{equation}
where $\Gamma(\cdot)$ is the complete Gamma function and $\Gamma(\cdot; \cdot)$ the incomplete Gamma function.
For values of $\gamma > 1$ we have a kurtosis smaller than 3, \ie probability densities, which
are less peaked than Gaussians. The effect of this parameter on the probability density can be seen in
Figs. (\ref{fig:suga12}.a) and (\ref{fig:suga12}.b). 

\begin{figure}
  \begin{center}
    \epsfxsize=0.49\textwidth
    \epsfbox{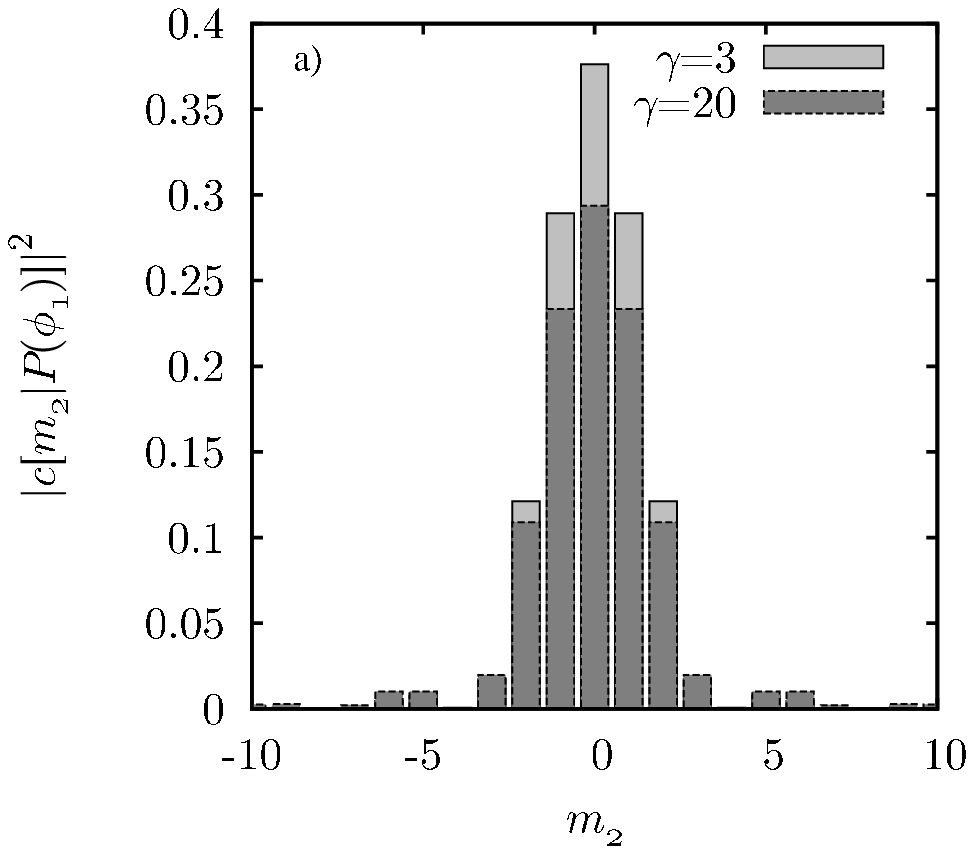}
    \epsfxsize=0.49\textwidth
    \epsfbox{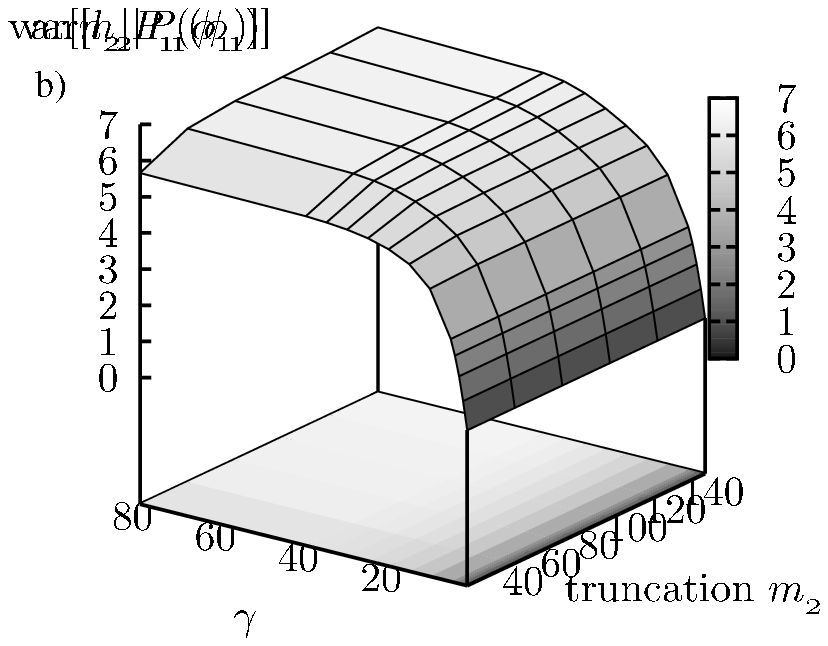}
  \end{center}
  \caption{\label{fig:suga34}$a)$ Numerically calculated OAM distribution for TSG apertures. The distribution is
  plotted for two different values of $\gamma$. In $b)$ the dependency of the conditional variance on the truncation
  index $m_2$ and $\gamma$ is shown. One can see that for small $\gamma$, \ie $\gamma <5$ the variance converges and stays
  independent of the truncation index. For higher $\gamma$ the variance converges more slowly, but there
  is no logarithmic divergence.}
  
\end{figure}

For the TSG all calculations have been done numerically and one can see in Fig (\ref{fig:suga34}.a) that, even 
for high values of $\gamma$, the probability distribution, $|c[m_2 | P_1(\phi_1)]|^2$, 
differs substantially from the rectangular case. For the variance, the dependency on the truncation index $m_2$
and $\gamma$ is shown in Fig. (\ref{fig:suga34}.b). For small values of $\gamma$ the variance does not change
over the range of truncation indices, for higher values the variance converges more slowly or not at all within
the given range. For the highest value of $\gamma=80$ the effect is more pronounced, but far from the
logarithmic increase in the rectangular case.

%----------------------------

%%% SUBSECTION: Discussion
\subsection{Discussion}
The theoretical modelling of a conditional angle measurement leads to a conditional variance significantly
different from zero. However, the final result depends not only on the aperture in the signal, which
sets the condition, but also on the aperture used to determine the angle. The angular EPR criterion Eq.
(\ref{eq:paracrit}) takes the orientation of $P_1(\phi_1;\tau_1)$ into account by averaging over
the the orientation angle $\tau_1$. From the simplified approach in section \ref{ssc:oam} we could assume
isotropic correlation for the azimuthal angle, which is expressed in the $\delta_{2\pi}$ function
in Eq. (\ref{eq:anglecorr}). Within this model the correlation are therefore the same for every
orientation $\tau_1$ and an averaging is not necessary. The dependence of $P[P_2(\phi_2) | P_1(\phi_1)]$
on the aperture functions $P_2$ and $P_1$ reflects the way in which the azimuthal angle is measured.

From the presented results, it is clear that the broadness
of the conditional angle probability density and the OAM probability distribution is mostly
determined by $P_1(\phi_1)$. The influence of the analyzing aperture $P_2(\phi_2)$ is certainly 
most controllable for the rectangular case. But the analysis of TSGs showed that there
are values of $\gamma$ for which the influence of $P_2(\phi_2)$ is comparable to the rectangular
case without showing the divergent variance. Using these apertures to study the influence of
the analyzing aperture on the conditional probability distribution could lead to a more 
detailed model for the measurement process and thus to a more complete theoretical prediction
for the final conditional OAM variance. On the other hand, for TSGs the question remains 
how well the aperture can be programmed in an SLM, in particular as for high values of $\gamma$ the 
features which distinguish it from a rectangular aperture are on a very small scale. This however might
be an interesting aspect in a more detailed analysis of rectangular apertures, which includes optical diffraction, 
as even small deviations from a perfect rectangular form would give a finite conditional variance.

%----------------------------

%%% SECTION: Conclusion
\section{Conclusion} 
In this work we have discussed the possibility to demonstrate an angular EPR paradox for the
conjugate variables of orbital angular momentum [OAM] and angle. The paradox is about the 
apparent violation of an uncertainty relation for incompatible observables measured on
correlated, spatially separated subsystems. By using photon pairs entangled in OAM and in angle these
subsystems can be realised in an optical experiment. We have found a testable criterion for an angular EPR paradox,
which takes experimental indeterminacies into account. For that we have reformulated 
the EPR paradox using conditional variances, \ie variances of observables from one subsystem 
given a preset outcome on the other subsystem.    

To investigate the feasibility of an experimental demonstration, we have modeled the measurement process under the 
assumption of perfect angle correlation. Also, angular apertures to set the condition or to determine
the angle were assumed to impose their probability characteristics exactly on transmitted photons.
Under these assumptions we have studied different classes of aperture functions for the final
conditional OAM variance. Rectangular functions lead to a divergent conditional OAM variance,
which does not set any lower bound for the correlations in OAM for our criterion. Truncated Gaussians
result in a quickly converging variance, but the implementation of the apertures with a smooth grayscale transition 
on an SLM will be less accurate than for rectangular apertures with their sharp edge. Truncated super Gaussians, which can be varied from 
the rectangular case to the truncated Gaussians, provide an aperture which leads to a convergent variance and still has
a controllable influence of the analysing apertures in the idler. The conditional variances obtained from the theoretical
modelling show that an angular EPR paradox can be demonstrated. Given the current state of experiments we expect an
implementation of our criterion to be able to demonstrate the EPR paradox for OAM and azimuthal position.

%%% SECTION: Acknowledgements
\section*{Acknowledgements}
We would like to thank Roberta Zambrini for useful discussions and for the suggestion to use truncated super Gaussians.
Also, Eric Yao and Miles Padgett have been a very helpful in questions regarding the experimental implementation. 
This work has received support from the Engineering and Physical Sciences Research Council [EPSRC] under the
grant GR S03898/01 and the Royal Society of Edinburgh.

%%% REFERENCES

\newcommand{\NAT}{\textit{Nature}}
\newcommand{\PR}{\textit{Phys. Rev.}}
\newcommand{\PRA}{\textit{Phys. Rev. A}}
\newcommand{\PRL}{\textit{Phys. Rev. Lett.}}
\newcommand{\NJP}{\textit{New Journal of Physics}}
\newcommand{\JMO}{\textit{J. Mod. Opt.}}
\newcommand{\JOB}{\textit{J. Opt. B}}

\end{document}